\documentclass[aps,prl,reprint,superscriptaddress,intlimits]{revtex4-2}
\usepackage{bm,latexsym,mathrsfs,enumerate,amsmath,amssymb}

\usepackage{graphicx}
\usepackage[breaklinks=true,urlcolor = blue,colorlinks = true,citecolor = blue,linkcolor = blue]{hyperref}
\usepackage[utf8]{inputenc}

\begin{document}

\title{Controlling Knot Topology in Magnetic Hopfions via Spin-orbit Torque}

\author{Shoya Kasai}
  \email{Contact author: kasai@aion.t.u-tokyo.ac.jp}
  \affiliation{Department of Applied Physics, The University of Tokyo, Hongo, Tokyo 113-8656, Japan}
\author{Shun Okumura}
  \affiliation{Department of Applied Physics, The University of Tokyo, Hongo, Tokyo 113-8656, Japan}
  \affiliation{Quantum-Phase Electronics Center (QPEC), The University of Tokyo, Hongo, Tokyo 113-8656, Japan}
  \affiliation{RIKEN Center for Emergent Matter Science (CEMS), Wako, Saitama 351-0198, Japan}
\author{Yukitoshi Motome}
  \email{Contact author: motome@ap.t.u-tokyo.ac.jp}
  \affiliation{Department of Applied Physics, The University of Tokyo, Hongo, Tokyo 113-8656, Japan}

\begin{abstract}
Knots, characterized by topological invariants called the Hopf number $H$, arise from the intertwining of strings and exhibit diverse configurations. The knot structures have recently been observed in condensed matters, as examplified by a magnetic hopfion, sparking interest in controlling their topology. Here, we show that spin-orbit torque (SOT) enables dynamic manipulation of the Hopf number of magnetic hopfions. We investigate the SOT-driven evolution of hopfions, revealing the splitting of a high-$H$ hopfion into multiple lower-$H$ ones, a process that can be quantified by an effective tension picture. Comparative analysis across different $H$ uncovers a hierarchy of instabilities that dictates these dynamical topological transitions. These findings establish SOT as a powerful tool for controlling hopfion topology, paving the way for potential applications in topological memory devices.
\end{abstract}

\maketitle

\begin{figure*}[t!]
  \centering
  \includegraphics[width=\hsize]{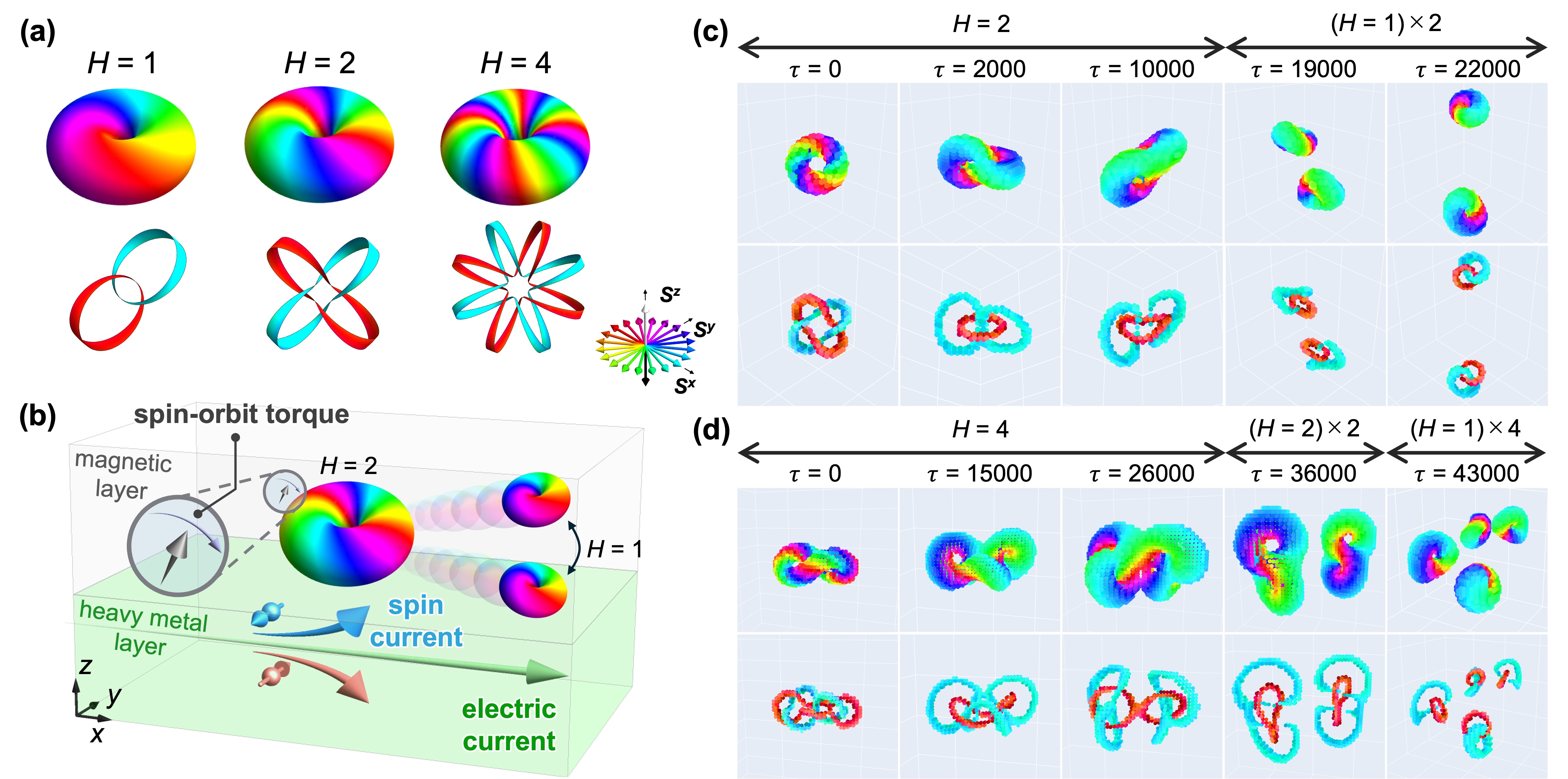}
  \caption{(a)~Schematic illustrations of magnetic hopfions with $H = 1, 2$, and $4$, shown by the isosurfaces of constant $S^z$, along with their corresponding preimages that satisfy $S^x = 1$~(blue) and $S^x = -1$~(red). The color code in the inset is used throughout this paper to depict spin orientations. (b)~A typical setup for this study. A spin current injected from the lower heavy metal layer works on magnetic moments as a SOT, thereby inducing hopfion dynamics in the upper magnetic layer. (c,d)~Snapshots of the $H = 2$ (c) and $H = 4$ (d) hopfions under $(B, \zeta) = (0.003, 0.002)$ and $(B, \zeta) = (0.003, 0.0021)$, respectively. The lower panels show their preimages to support the observation of changes in the knot topology.}
  \label{schematic}
\end{figure*}

Knots, formed by the intertwining of one or more strings, have a great variety of topologically different forms that cannot be transformed into one another by continuous deformation. Recognizing this diversity, Lord Kelvin proposed the vortex theory of atoms, where each chemical element was associated with a knot of vortex filaments in the ether~\cite{Thomson1869}. Although the theory has since been abandoned, it inspired the development of knot theory, which classifies topologically different knots by studying knot invariants~\cite{Adams2004}. The knot theory now plays a crucial role in various scientific disciplines, including string theory~\cite{Zwiebach2009}, quantum field theory~\cite{Witten1989}, molecular machines~\cite{DietrichBuchecker1989}, DNA topology~\cite{Bates2005}, and topological quantum computing~\cite{Kitaev2003}.

When it comes to knots formed by two strings, a useful approach for distinguishing topologically different knots is to compute the linking number, also known as the Hopf number $H$, which represents the number of times the strings are linked. $H$ is an integer characterizing the homotopy group $\pi_3(S_2) \cong \mathbb{Z}$ that classifies continuous maps from the three-dimensional (3D) sphere $S^3$ to the two-dimensional (2D) sphere $S^2$~\cite{Hopf1931}. Physical realizations of this topology have been discovered in various systems such as high-energy physics~\cite{Faddeev1997}, optics~\cite{Sugic2021,Shen2023}, superfluid $^{3}$He~\cite{Volovik1977}, triplet superconductors~\cite{Babaev2002}, bosonic systems~\cite{Kawaguchi2008,Hall2016}, ferroelectrics~\cite{Lukyanchuk2020}, and liquid crystals~\cite{Ackerman2017_nmat,Ackerman2017_PRX,Tai2019}, where certain linked structures can be found. Moreover, stimulated by the recent direct observations of magnetic hopfions~\cite{Kent2021,Yu2023,Zheng2023}, which are the magnetic counterparts of the knot structures, increasing attention has been directed toward novel quantum phenomena originating from the knot topology.

Magnetic hopfions are 3D topological spin structures with the knot topology. Figure~\ref{schematic}(a) displays the isosurfaces defined by constant $S^z$, $z$ components of spins, for hopfions with $H = 1, 2,$ and $4$, which resemble tori. Each lower inset shows a preimage, a closed loop that satisfies $S^z = 0$ and $S^x = \pm 1$. Any pair of the preimages is linked an integer number of times, which corresponds to the Hopf number $H$. While the stability of the hopfions was initially demonstrated in confined geometries where the sample shape anisotropy plays a crucial role~\cite{Sutcliffe2018,Liu2018,Tai2018,Li2022}, recent studies have actively explored their stability and dynamics in bulk systems with competing interactions~\cite{Bogolubsky1988,Sutcliffe2017_frustration,Rybakov2022,Sallermann2023,Lobanov2023,Liu2020,Liu2022}. A particularly intriguing fact is that hopfions can, in principle, be stabilized with arbitrary Hopf numbers~\cite{Rybakov2022}. Furthermore, reflecting the characteristic energy scaling with Hopf number~\cite{Vakulenko1979}, multiple hopfions tend to merge into a single hopfion whose Hopf number equals the sum of the originals~\cite{Ward2000,Hietarinta2012,Kasai2025}. This distinctive property poses a fundamental challenge for controlling magnetic hopfion topology: how can a high-$H$ hopfion be split into lower-$H$ ones? While such reconfiguration of knot topology has recently been proposed in chiral liquid crystals using pulse electric fields~\cite{Hall2025}, a method for inducing analogous dynamical transitions in magnetic hopfions has yet to be known.

In this Letter, we demonstrate that spin-orbit torque (SOT) can induce the splitting of a high-$H$ hopfion into multiple lower-$H$ hopfions, thereby enabling control over their topological property, through numerical solutions of the Landau-Lifshitz-Gilbert (LLG) equation. First, we find that the SOT triggers the splitting of the hopfions with $H = 2$ and $4$, eventually leaving multiple hopfions with lower Hopf numbers. Next, we introduce an ``effective tension" picture, which enables both qualitative and quantitative analyses for the splitting process. Finally, by comparing the SOT-driven dynamics across different $H$, we show that the splitting originates from a hierarchichy associated with the stability of the hopfion with each Hopf number under the SOT. Our results manifest the rich tunability of the knot topology of magnetic hopfions and pave the way for the development of multilevel memory devices that exploit the switching performance.

We investigate the hopfion dynamics driven by the SOT known as a technique for magnetization switching~\cite{Miron2011,Liu2012}. A typical setup is shown in Fig.~\ref{schematic}(b), where an electric current (green arrow) in the lower heavy metal layer with strong spin-orbit coupling generates a spin current in a perpendicular direction (blue and orange arrows), injected into the upper magnetic layer. The spin current exerts the SOT on the magnetic moments, leading to the dynamics of hopfions in the upper magnetic layer. For the magnetic layer, we consider a spin model with competing interactions on a simple cubic lattice. The Hamiltonian is given by
\begin{align}
\mathcal{H} = -\sum_{\alpha = 1}^4 \sum_{\langle i,j \rangle_{\alpha}} J_{\alpha}~\mathbf{S}_i \cdot \mathbf{S}_j - B \sum_{i} S_i^z,
\label{eq:model}
\end{align}
where $\mathbf{S}_i = (S_i^x,S_i^y,S_i^z)$ represents the classical spin at site $i$ with $|\mathbf{S}_i|=1$. The first term represents the exchange interactions; the summation with respect to $\langle i,j \rangle_{\alpha}$ runs over $\alpha$th-neighbor pairs on the cubic lattice. We take $(J_1,J_2,J_3,J_4) = (1,-0.166,0,-0.083)$, and note that similar parameters were employed to stabilize a hopfion~\cite{Bogolubsky1988,Rybakov2022}. The second term describes the Zeeman coupling to the magnetic field $B$. We set the lattice constant as unity and adopt periodic boundary conditions. For an analysis of the hopfion dynamics driven by the spin current, we numerically solve the LLG equation with the SOT, which is given as
\begin{align}
  \frac{d \mathbf{S}_i}{d\tau} = \mathbf{S}_i \times \mathbf{H}_i^{\rm eff} + \alpha \mathbf{S}_i \times \frac{d \mathbf{S}_i}{d\tau} + \zeta \mathbf{S}_i \times (\mathbf{S}_i \times \mathbf{p}),
  \label{eq:LLG}
\end{align}
where $\tau$ denotes dimensionless time, and $\alpha$, $\zeta$, and $\mathbf{p}$ represent the Gilbert damping, the strength of the SOT, and the polarization direction of the spin current, respectively. The effective field $\mathbf{H}_i^{\rm eff}$ in Eq.~\eqref{eq:LLG} is defined as $\mathbf{H}_i^{\rm eff} = \frac{\partial \mathcal{H}}{\partial \mathbf{S}_i}$. To solve Eq.~\eqref{eq:LLG} we use the fourth-order Runge-Kutta method. We take the Gilbert damping $\alpha = 0.2$, the time step $\Delta \tau = 0.1$, and $\mathbf{p} = \mathbf{\hat{y}}$. We prepare an initial spin state including hopfions for the LLG simulations by performing a gradient-based energy minimization for the spin configuration based on the ansatz in Ref.~\cite{Hietarinta1999,Wang2019}. For details of the computational procedure, refer to the previous work~\cite{Kasai2025} and our supplemental submission.

First, we investigate the SOT-driven dynamics of an $H = 2$ hopfion. Figure~\ref{schematic}(c) displays hopfion snapshots of the dynamics under $(B, \zeta) = (0.003, 0.002)$. The colored dots represent spin vectors visualized according to the color code shown in Fig.~\ref{schematic}(a). The leftmost panels show the initial hopfion and its preimages prepared by the above procedure. These structures are equivalent to those in the middle panel of Fig.~\ref{schematic}(a). As shown in the panel at time $\tau=2000$, the SOT works to twist and distort the torus shape of the hopfion, which in turn significantly alters the shape of the preimages. While the red preimage maintains a single closed loop, the blue preimage transforms into an ``8"-shaped form. Subsequently, at $\tau = 10000$, the blue preimage begins to split into two rings from the junction in ``8". This can be regarded as a precursor to a transition of the knot topology of the hopfion. Indeed, as shown in the panel at $\tau=19000$, the $H=2$ hopfion splits into two $H=1$ hopfions. Their preimages consist of red and blue loops linked once, topologically equivalent to the preimages in the left panel of Fig.~\ref{schematic}(a).

Next, we examine the SOT-driven dynamics of an $H = 4$ hopfion. Figure~\ref{schematic}(d) displays snapshots of the dynamics under $(B, \zeta) = (0.003, 0.0021)$. In the current model in Eq.~\eqref{eq:model}, it is known that hopfions with $H \geq 3$ tend to stabilize with lower-symmetric isosurfaces compared to the torus shapes shown in Fig.~\ref{schematic}(a). The leftmost panel displays an energetically optimized $H=4$ hopfion, whose ``8"-shaped isosurface is consistent with a previous study~\cite{Rybakov2022}. As shown in the snapshot at $\tau = 15000$, the SOT induces a complex deformation of the hopfion and its preimages, and eventually the cyan preimage splits into two by $\tau = 26000$. The two split cyan preimages are barely connected by the red preimage which has an ``8" shape. Upon further time evolution, this red junction is disconnected and two independent $H=2$ hopfions remain, as shown in the snapshot at $\tau = 36000$. Each preimage is topologically the same as that in Fig.~\ref{schematic}(c) at $\tau = 10000$. Subsequently, these two $H=2$ hopfions exhibit dynamics similar to those observed in Fig.~\ref{schematic}(c), eventually breaking down into four individual $H=1$ hopfions. This multistep topological transitions from an $H = 4$ state to an $H = 1 \times 4$ state via an $H = 2 \times 2$ state are owing to the remarkable diversity of the knot topology of hopfions. This stands in contrast to 2D skyrmions because the skyrmions with topological charge greater than $4$ are unstable, making it difficult to study their dynamics~\cite{Xichao2017}.

\begin{figure}[t!]
  \centering
  \includegraphics[width=\hsize]{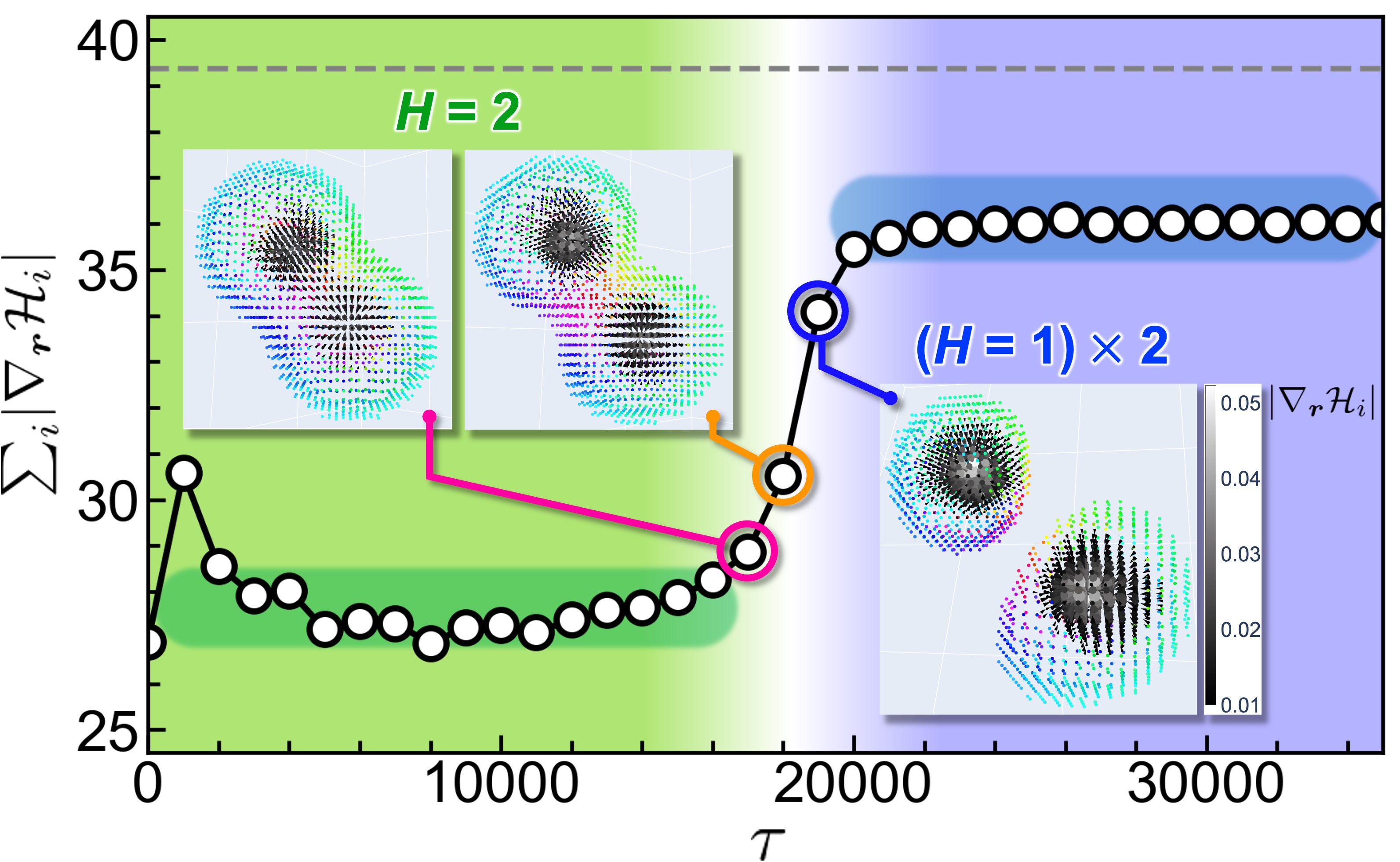}
  \caption{Time evolution of the absolute sum of the spatial gradient of energy for the result of Fig.~\ref{schematic}(c). Splitting occurs around the white time region, during which the value rapidly increases. The horizontal gray dashed line is the reference corresponding to twice the value of the gradient sum of a relaxed $H=1$ hopfion in the absence of the SOT. The insets shows the dynamics around the splitting. The colored dots represent the spins at each site. The arrows shown in grayscale represent the gradient vectors $\nabla_{\mathbf{r}} \mathcal{H}_i$ and their norms.}
  \label{H2_tension}
\end{figure}

Let us discuss the mechanism responsible for the splitting dynamics, focusing on the $H=2$ case. The twisted and stretched structures at $\tau = 2000$ and $\tau = 10000$ in Fig.~\ref{schematic}(c) suggest that the SOT works as an ``effective tension" on the hopfion, pulling it in two opposite directions. When a sufficiently strong SOT is applied, it breaks the topological protection derived from the knot topology and tears the hopfion apart. To examine this picture, we introduce the absolute sum of the spatial derivatives of the energy as an indicator to quantify the effective tension, which is given by $\sum_i |\mathbf{\nabla}_{\mathbf{r}} \mathcal{H}_i|$ with the internal energy $\mathcal{H}_i$ at site $i$. Figure~\ref{H2_tension} shows the value as a function of time $\tau$ for the dynamics in Fig.~\ref{schematic}(c). In the green-shaded region before the splitting at $\tau \simeq 19000$, the value remains around 27. In contrast, it rises sharply toward the splitting and converges to approximately $36$ in the blue-shaded region after the splitting, which is close to the value $\simeq 39$ for two independent $H=1$ hopfions indicated by the gray dashed line. The insets show the spatial distributions of ``effective tension vector" $\nabla_{\mathbf{r}} \mathcal{H}_i$ (grayscale arrows), along with the spin structures (colored dots). Before the splitting, the effective tension is broadly and thinly distributed over the stretched structure, whereas after the splitting it becomes localized around the core of each hopfion with $H = 1$. The surge in the spatial derivative and its localization can be regarded as the ``work" stored in the hopfions, done by the SOT as an external effective tension. We performed a similar analysis for an $H = 4$ hopfion and found that the effective tension picture provides an efficient means of understanding the multistep topological transitions in the knot topology shown in Fig.~\ref{schematic}(d); see the supplemental submission.

\begin{figure}[t!]
  \centering
  \includegraphics[width=\hsize]{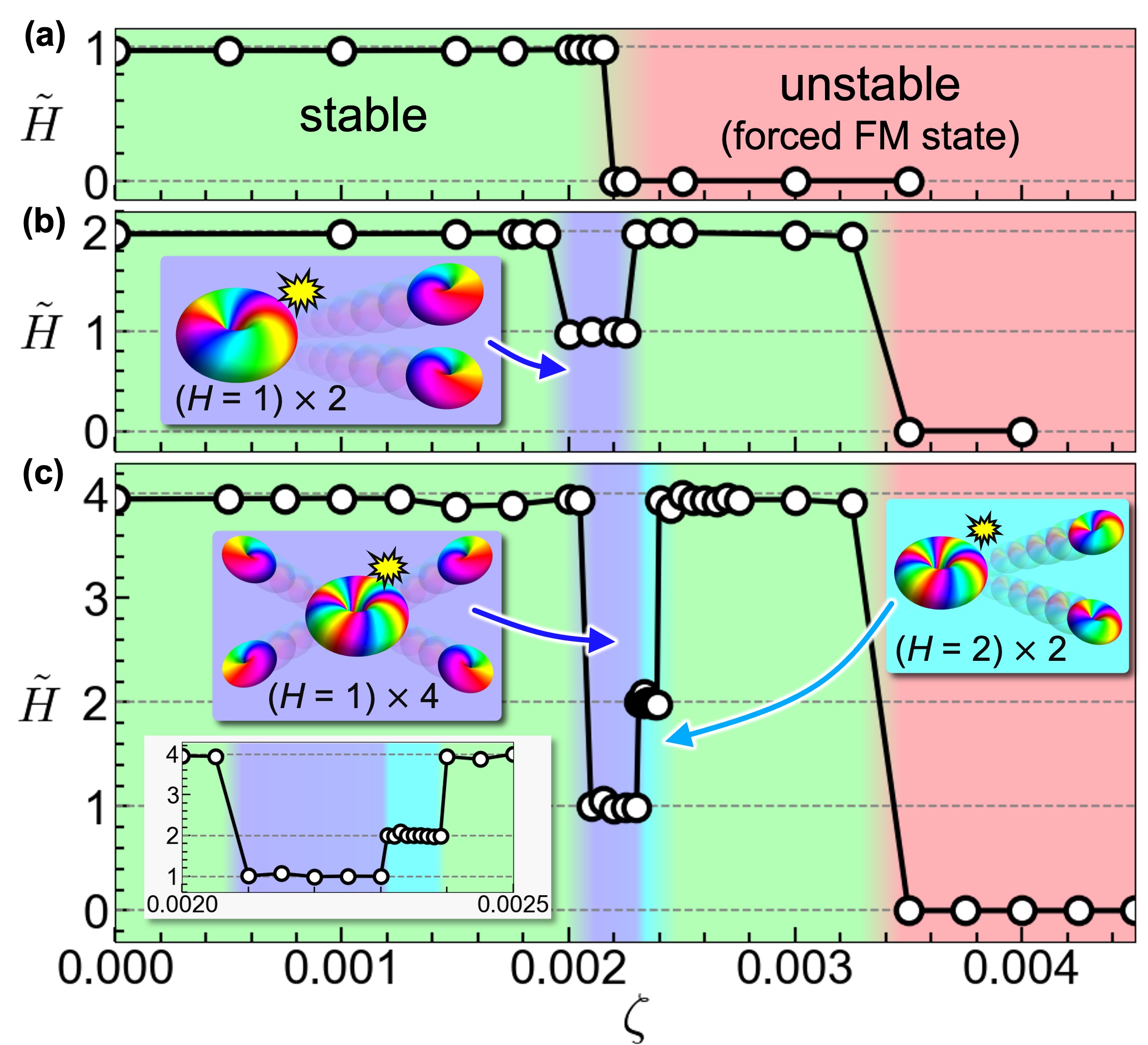}
  \caption{SOT-driven steady-state phase diagrams for (a)~$H = 1$, (b)~$H = 2$, and (c)~$H = 4$ hopfions, with plots of the local Hopf number $\tilde{H}$ as a function of $\zeta$. The lowerleft inset of (c) shows an enlarged plot for $0.0020 \lesssim \zeta \lesssim 0.0025$. We take $B=0.003$.}
  \label{H2_1Ddiagram}
\end{figure}

Given the effective tension picture of the SOT, we perform systematic simulations to clarify the range of the SOT strength $\zeta$ required for the splitting dynamics. To detect the splitting, we introduce a local Hopf number $\tilde{H}$, computed over a finite cubic region centered around the hopfion core. The cube size is chosen to closely approximate the value of $H$ for each isolated hopfion. For example, in the time evolution in Fig.~\ref{schematic}(c), $\tilde{H}$ decreases from $2$ to $1$ because the cubic region for computing $\tilde{H}$ encompasses only one of two $H = 1$ hopfions after the splitting. Thus, the value of $\tilde{H}$ after sufficient time evolution is useful to discuss the hopfion state in the nonequilibrium steady state. Figure~\ref{H2_1Ddiagram} summarizes the $\zeta$ dependence of $\tilde{H}$ in the steady state starting from (a) $H=1$, (b) $H=2$, and (c) $H=4$ states.

Let us first discuss the $H=2$ case in Fig.~\ref{H2_1Ddiagram}(b). $\tilde{H}$ reaches $1$ at an intermediate SOT strength in the purple-shaded region, indicating that the $H=2$ hopfion splits into two $H=1$ hopfions within this regime, as shown in Fig.~\ref{schematic}(c). In contrast, in the green-shaded regions on both sides, $\tilde{H}$ remains at $2$, signifying that no splitting occurs. Despite the common absence of splitting, the $H=2$ hopfion behaves differently in these two regions. In the left green-shaded region with weak SOT, the hopfion deforms into an elliptical shape but shows no spatial motion, whereas in the right green-shaded region with strong SOT, it exhibits both translational and rotational motion; see the supplemental submission for details. Finally, in the red-shaded region, excessively strong SOT destabilizes the hopfion, resulting in a topologically trivial state with an in-plane ferromagnetic polarization.

Next, we turn to the $H=4$ case shown in Fig.~\ref{H2_1Ddiagram}(c). In this case, in addition to the splitting into four $H=1$ hopfions as shown in Fig.~\ref{schematic}(d) at an intermediate SOT in the purple-shaded region, we observe an additional steady state shaded in cyan, adjacent to that regime. In this region, as clearly shown in the inset, the $H = 4$ hopfion splits into two $H = 2$ hopfions, similar to what is shown in the middle-right panel of Fig.~\ref{schematic}(d), but does not subsequently split into four $H = 1$ hopfions within our simulations. We note that in this regime the dynamics depends on the initial states; we plot representative results from simulations initiated with 15 distinct configurations, carefully examining the values of $\tilde{H}$ and corresponding snapshots. In the other green and red-shaded regions, the steady states are similar to those in the $H=2$ case in Fig.~\ref{H2_1Ddiagram}(b).

The similarity between the steady-state phase diagrams for the $H=2$ and $H=4$ cases, except for the cyan region unique to the latter, suggests a potential relationship between them. To explore this further, we include the phase diagram for the $H=1$ case in Fig.~\ref{H2_1Ddiagram}(a). In Fig.~\ref{H2_1Ddiagram}(a), the region where the hopfion remains stable is shaded in green. In this region, the hopfion exhibits both translational and rotational motions (see the supplemental submission for details). In contrast, in the red-shaded region for $\zeta > \zeta_{\rm c} \simeq 0.0022$, the hopfion becomes unstable and results in a trivial forced ferromagnetic state. Importantly, as shown in Fig.~\ref{H2_1Ddiagram}(b), the critical value of $\zeta$ between the purple-shaded and the right green-shaded regimes coincides with $\zeta_{\rm c}$ for $H = 1$. This suggests that the higher-$H$ hopfion undergoes splitting under sufficiently strong SOT, but ceases to split in the regime where the lower-$H$ hopfion becomes unstable. Consistently, the value of $\zeta$ at which the $H = 4$ hopfion ceases to split into four $H = 1$ hopfions also approximately coincides with $\zeta_{\rm c}$, and it enters the cyan-shaded region where $H = 4$ is able to split into two $H = 2$. This behavior is also understood from the lower-$H$ steady states: above the threshold, the $H=1$ hopfion is unstable, whereas the $H=2$ hopfion remains stable without splitting. These analyses reveal a hierarchy among the steady-state phase diagrams for different $H$, where the behavior of higher-$H$ hopfions is constrained by the stability of lower-$H$ hopfions. Given this hierarchy, we can predict the SOT-driven dynamics of hopfions with other values of $H$ without performing additional simulations, as detailed in our supplemental submission.

To conclude, we have investigated the stability and real-time dynamics of high-$H$ hopfions under the SOT. Starting with an $H=2$ hopfion, we discovered that it splits into two $H=1$ hopfions under sufficiently strong SOT. In addition, intriguingly, we found that an $H=4$ hopfion has two splitting regimes depending on the SOT strength: one where it splits into four $H=1$ hopfions and another where it splits into two $H=2$ hopfions. We also analyzed these dynamical topological transitions using the effective tension picture. Furthermore, we revealed a hierarchy in the instability toward splitting across different Hopf numbers. These rich controllability of knot topology is a unique property to the hopfions which can be stabilized with arbitrary Hopf numbers. Our findings, combined with their intrinsic capability for spontaneous fusion, pave the way for potential applications in innovative magnetic memory devices, such as multilevel memories, that harness the rich topological degrees of freedom of hopfions and their mutual convertibility under the SOT.

We thank M. Ezawa, Y. Kato, and K. Shimizu for fruitful discussions. This work was supported by the JSPS KAKENHI (No.~JP22K13998, JP23K25816, and JP25H01247) and JST PRESTO (No.~JPMJPR2595). S. K. was supported by the Program for Leading Graduate Schools (MERIT-WINGS) and JST SPRING, Grant Number JPMJSP2108. The computation in this work has been done using the facilities of the Supercomputer Center, the Institute for Solid State Physics, The University of Tokyo.



\bibliography{bibliography}
\end{document}